# SEQUENCING AND NAVIGATION THROUGH LEARNING CONTENT

*Gordana Rakić, Ljubomir Jerinić, Zoran Budimac, Mirjana Ivanović*
Department for Mathematics and Informatics
Faculty of Sciences
University of Novi Sad
Trg Dositeja Obradovića 4, 21000 Novi Sad, Serbia
Tel: +381 21 4852877; fax: +381 21 6350458
e-mail: {goca, jerinic, zjb, mira}@dmi.uns.ac.rs

**ABSTRACT**

The aim of this paper is to describe a basic idea to introduce formal modeling in development and presentation of e-learning systems. By introducing formalism, reasoning on e-learning systems and processes through them can be more easily understood. In particular, we propose to use Harel's automata (statecharts) in modeling sequencing and navigation through learning content and learning process.

## 1 INTRODUCTION

Nowadays, e-learning play important role in educational market. Learning systems with confident management of learning contents and process bring additional value to the e-learning services. Formalization and pre-formalization in this area, that would provide more confidence in learning outcome, could assure additional quality to the learning process and results.

The design is drawn from the model OSOF [5, 7] where authors introduce full terminology and structure for the representation of different kind of information, which could be used in learning by the aims of computer, with the intention to introduce navigation trough that kind of information. Proposed approach is adapted to support SCORM standard but idea remains the same.

Most modern e-learning systems are based on the collection of various kind of information that needs to be accepted by the student, the presentation of this information on the computer, and the methods of their use by students. This approach assumes that the teaching material is presented as the mathematical structure of the graph. However, learning is primarily managed process by teachers who decide whether a material is adopted or not, as well as how to help the student when certain problems occur in the acquirement of content. Therefore, any information that a student needs to pass (overcome) has a beginning, and under certain conditions, the student moves on to the next. Therefore, the mathematical aspects of the structure of a series of information that a student needs to pass (overcome) has to be in a form of finite automata, in our case modified *Harel's automata*.

In this paper we propose formal modeling of learning content and process. In Section 2 the brief overview of course management systems is provided. E-learning terminology recommended by SCORM standard for e-learning interoperability is provided in section 3, while statecharts and their usage in formal modeling are introduced in section 4. Idea for statechart modeling of e-learning content and process is described in section 5. Related work is described in section 6, while conclusion and future work are given in section 7.

## 2 COURSE MANAGEMENT SYSTEMS

Course organization is an important aspect in educational process. Digitalization and usage of electronic courses gives additional value to good course management.

There are numerous approaches to course management which can be divided in two basic categories: course management systems oriented to learning process and course management systems oriented to learning content.

Systems oriented to learning content are aimed to organization of learning material independently on its usage in learning process. Usually, for these purposes we use Course Repositories (CR) or Learning Content Management Systems (LCMS). The main difference between these two categories are that CR do not take care about creation of material, but only about storing and delivery of stored content.

Systems oriented to learning process as a goal have management of creating, storing, organizing, and using learning material in learning process. There are two categories of process oriented learning systems: Learning Management Systems (LMS) and Intelligent Tutoring Systems (ITS). ITS uses techniques of artificial intelligence to adapt learning process and content to learner and to provide learner with instructions and feedback during the learning process.

Intelligent or not, process oriented learning systems include learning strategies to provide adaptive learning. In this context by learning strategy we mean personalized approach to learner but with methodology, pedagogy and psychology driven teaching rules. Therefore we could equally use the term "teaching strategy", but we will keep the "learning strategy" as a common phrase.. Adaptability gives to these systems dimension of reactivity.

## 3 SCORM

Sharable Content Object Reference Model (SCORM)[1] is a set of technical standards for e-learning interoperability. It is developed in year 2000 by US Department of Defense organization called Advanced Distributed Learning (ADL)[2] Nowadays SCORM is widely used to transfer courses across different LMS and LCMS. It contains needed standards, specifications and guidelines for describing the relationship of content objects, data models and protocols such that objects become sharable across the systems that meet this standard.

SCORM standard consists of the following three parts:
- Run-Time Environment (SCORM RTE) takes care about launching of the content by LMS and way the content communicates with the LMS.
- Content Aggregation Model (SCORM CAM) defines learning content and its organization. SCORM defines two levels of contents: a) asset and b) Sharable Content Object (SCO). Asset is any piece of digital information (text, image, sound, etc.), while SCO is a collection of one or more assets that represent a single launchable learning resource. SCO is a smallest unit of information. (it can communicate with LMS by utilization of SCORM RTE.
- Sequencing and Navigation (SCORM SN) allows the course maker to govern how the learner is allowed to navigate between SCOs. It is based on hierarchy tree of activities represented by items. Item can be either asset or SCO if item is leaf in the tree. Otherwise it is called cluster and contains child activities. The rules for order in which user will be passing through the units (SCO) is realized by sequencing. Navigation provides possibility for learner to follow specific flow through the contents. This term is used for process of movement through provided contents. By usage of different sequencing and navigations we can introduce learning strategies.

## 4 STATECHARTS IN FORMAL MODELING

Modeling plays important role in software development, especially when we are dealing with reactive systems. It provides problem abstractions and structure for problem solving, enables complexity management and experimentation in order to explore multiple solutions and to select the most appropriate one. Modeling helps in reduction of time-to-market for business problem solutions and development costs, but also in managing the risk of mistake. Formal modeling includes mathematical background as an additional value.

Visual problem representation provides better understandability of problem and its solution. Testability of some visual models is important feature in this direction. Therefore visual modeling is used to reveal possible gaps in software development.

Statecharts (or Harel's Automata) [2] provide all listed benefits. This is powerful technique for visual formal modeling which belongs to state based modeling approaches. Similarly to finite state automata, statechart model is usually runnable, and hence testable. Still, in comparison with finite state automata it offers some improvements, out of which we emphasize the following ones:
- multi-states and state hierarchy modeling,
- parallel states,
- time modeling.

Our plan is to use statechart to model adaptive learning system.

## 5 STATECHART MODEL FOR SEQUENCING AND NAVIGATION

If we observe learning system and its usage in learning process we can notice that at each moment single learner, and therefore from this perspective the whole system can be in the single state. This state is uniquely determinable. It is determined by current item. Item can be SCO or Asset (regular or assessment one). Asset is atomic, while SCO can consist of Assets. On the other hand, when system is stated in the certain item it is hierarchically uniquely stated in all higher levels of organization: topic, lesson, section, course and curriculum.

Navigation through the contents and the processes is based on events appeared in the process. We provide (Figure 1) only the basic idea for modeling of these events in the learning process described on low level. This idea could be propagated on higher levels with minimal modifications.

Let us observe Item. In the current lesson and in the current topic user can use single SCO that is defined as set of Assets. Therefore we define it recursively. SCO built by one Asset and other SCO. In theory this differs a bit from the SCORM definition of SCO, but in practice it fulfills the requirement.

Based on parameters affecting the event on entering the SCO system user enters to the Asset, Assessment Asset or to empty asset in which case we move to the exit point of the SCO. When we are on the exit point of the SCO we can come to one of the following situations:
- There is next content in current Item but learner did not pass assessment and system move learner back to the previous step (usually beginning of the Item dependently on strategy)

---

[1] Sharable Content Object Reference Model (SCORM)
http://scorm.com/
[2] Advanced Distributed Learning (ADL)[2]
http://www.adlnet.gov/

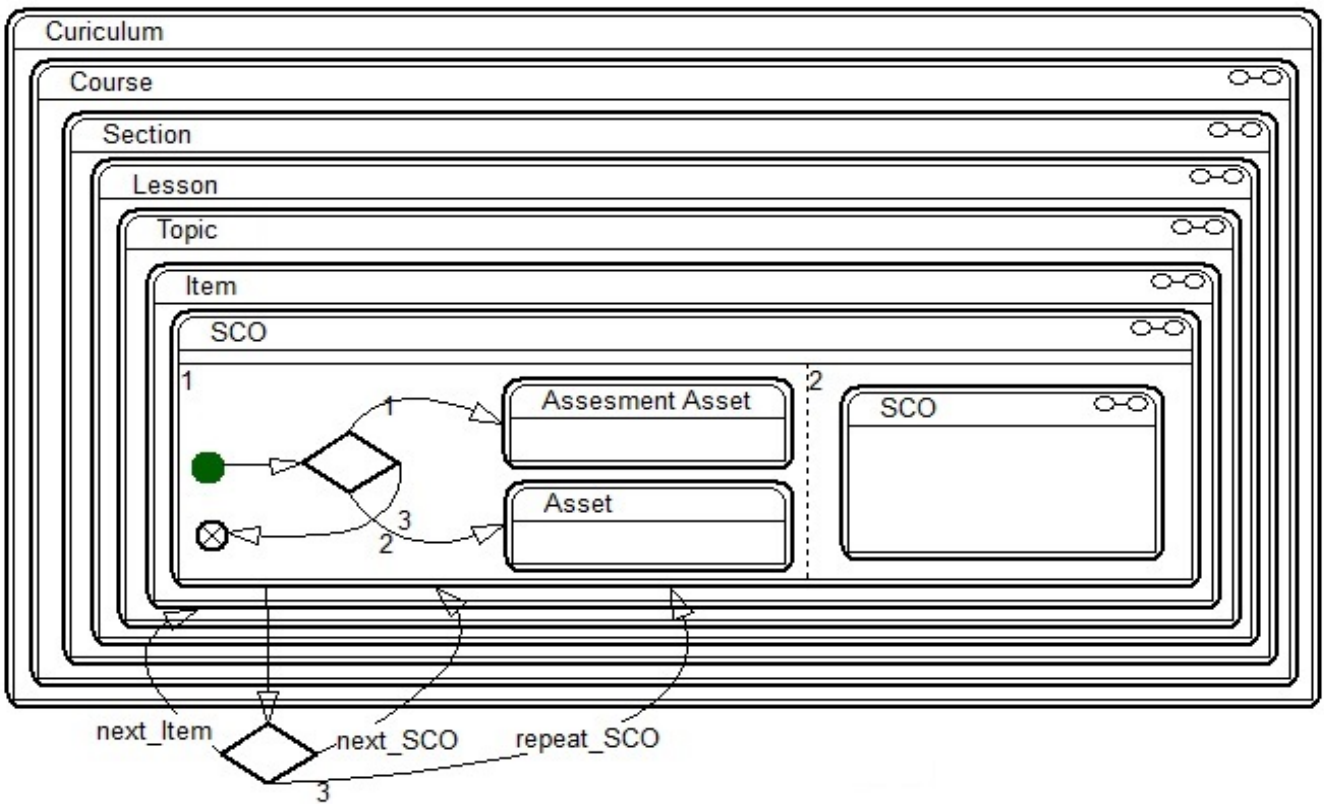

**Figure 1.** Statechart model for e-learning sequencing and navigation

- There exists next content in current Item and learner passed assessment and system move learner to the next content in the Item
- There is no next content in the Item and system moves the user to the next Item.

If we propagate these rules to the higher levels of the learning organization we will get full navigation through the curriculum.

Basic idea is to adapt conditions and events based on learning strategies.

## 6  RELATED WORK

Formal methods are commonly used in modeling critical properties of reactive systems. If we observe adaptive learning systems as a mission critical reactive system where the mission is successful fulfillment of learning goals, critical point is related to sequencing and navigation and the way for reaching the outcome. Sequencing and navigation are responsible to ensure learner to overtake all learning steps and to pass all needed tests. Formal modeling of sequencing should provide confidence in learning outcomes. Formal modeling in adaptive sequencing and navigation should provide confidence of learning strategy. Authors of [1] provide good overview with future trends in sequencing modeling. In a frame of UML based sequencing, UML state diagrams, as a version of statecharts are used for visual representation of navigation rules and states. One model is also provided as an example, but without any relation to available standards in the field. In graph-based modeling finite state automata are considered as an appropriate technique. In our context, statecharts are used as an extension on finite state automata integrating both approaches, but additionally following the SCORM standard, which was not case in provided overview. Furthermore, the overview is concentrated only on sequencing while our goal is to integrate the sequencing and the navigation in the formal model.

Authors of [6] provide conceptual meta-model for the educational application. Even if their goal is similar to ours there exist some important differences. The most important one is that authors of proposed paper provide the meta-model as a guidelines or pattern for further conceptual development of educational applications. Our goal is to develop the formal model of concrete learning system.

In [4] we can find model-based approach but only to SCORM sequencing oriented to modeling content organization, while our goal is to include also the navigation and to model the process.

Finally, authors of [3] take into consideration the learning process, but again the goal is to model learning content organization, not the learning process which is goal of our efforts.

# 7 CONCLUSION AND FUTURE WORK

This paper describes a brief idea for formal modeling of learning content and learning process by statecharts. The goal is to use formal model in order to meet dependability of learning systems.

By formally modeling learning contents as an automaton, we get the possibility to represent personalized learning as function P that would be applied to existing automaton, giving a new, personalized (or otherwise changed) one. Those functions could be also composed in different ways, thus providing different pedagogical flavors. Described idea for modeling of the navigation is demonstrated on the lowest level of sequencing, and possibility for propagation to the higher levels is described.

Still there are many open questions for investigation. All details are to be modeled and model is to be tested.

In this observation situations of following more then one course in the same curriculum in parallel, or even following more then one curriculums in parallel are not considered.

The real future work is to add dynamics to the model, by involving intelligent techniques for realtime generation of events and conditions for switching the states based on learning strategy and previous activities and results of the learner.

## ACKNOWLEDGMENT

The work descried in this paper was partially supported by the Serbian Ministry of Education, Science and Technological Development through project "Intelligent Techniques and Their Integration into Wide-Spectrum Decision Support," no. OI174023.